# RANDOM SEQUENTIAL ADSORPTION ON A LINE:
# MEAN-FIELD THEORY OF DIFFUSIONAL RELAXATION

**Vladimir Privman**[a,b] and **Mustansir Barma**[a,c]

[a] *Department of Physics, Theoretical Physics, University of Oxford,*
    *1 Keble Road, Oxford OX1 3NP, UK*

[b] on leave of absence from *Department of Physics, Clarkson University,*
    *Potsdam, New York 13699–5820, USA*

[c] on leave of absence from *Tata Institute of Fundamental Research,*
    *Homi Bhabha Road, Bombay 400005, INDIA*

## ABSTRACT

We develop a new fast-diffusion approximation for the kinetics of deposition of extended objects on a linear substrate, accompanied by diffusional relaxation. This new approximation plays the role of the mean-field theory for such processes and is valid over a significantly larger range than an earlier variant, which was based on a mapping to chemical reactions. In particular, continuum-limit off-lattice deposition is described naturally within our approximation. The criteria for the applicability of the mean-field theory are derived. While deposition of dimers, and marginally, trimers, is affected by fluctuations, we find that the $k$-mer deposition kinetics is asymptotically mean-field like for all $k = 4, 5, \ldots, \infty$, where the limit $k \to \infty$, when properly defined, describes deposition-diffusion kinetics in the continuum.

# 1. INTRODUCTION AND DEFINITIONS

Recent experimental advances [1,2] in the studies of kinetics of formation of amorphous layers of monodispersed submicron colloid particles and proteins on flat surfaces have stimulated much theoretical interest in models of irreversible growth [2]. Monolayer growth by random sequential adsorption was the earliest such system to be considered [2,3], and many theoretical results have been accumulated over the years. More recent efforts have been focused on multilayer structures, on results which can be obtained by modern computer simulations, and on allowing for relaxation in the deposit [4,5]. The subject of this work is diffusional relaxation in monolayer deposition.

In colloid deposition on surfaces, relaxation is typically slow. Diffusional relaxation can, however, be observed experimentally in protein monolayers and multilayers [6]. Furthermore, multiunit-long, "sliding" molecules can be attached to DNA suggesting consideration of one-dimensional lattice and continuum models. Related models of deposition of the epitaxial type (i.e., crystalline-structure forming), accompanied by diffusional and other relaxation processes, were considered in the literature [7].

Theoretical study of the irreversible adsorption with diffusion encounters several difficulties. Firstly, numerical simulations have proved exceedingly resource-consuming already even for one-dimensional models [4,5]. Secondly, even in the absence of deposition, systems of diffusing hard-core extended objects are not fully understood in dimensions higher than one, and are the subject of much current research and controversy [8]. Indeed, both the dynamics and equilibrium state of hard-core systems show complicated cooperative effects such as slow "glassy" relaxation, phase transitions and phase separation, etc.

Results of numerical simulations of lattice models in $1d$ were interpreted [5] within a certain mean-field type framework which will be described shortly. First, we have to introduce some notation. Consider deposition of extended objects of size $\ell$ on a line, with the attempt rate $R$ (per unit length and time). The attempts are uniformly distributed, but only those are successful which fall in the space free from objects deposited earlier. Initially, the substrate (1$d$ line) is empty. The deposited objects hop randomly, with the single-object (dilute-limit) diffusion constant $D$. Lattice models of $k$-mer deposition are defined by further introducing the space mesh in steps of

$$b = \ell/k , \qquad (1.1)$$

and requiring that both the deposition events and the hopping keep the extended objects aligned and registered with the resulting 1$d$ lattice. For instance, the deposition attempt rate is $Rb$ per lattice site (per unit time).

The limit $D = 0$ corresponds to deposition without diffusion, and provided $k > 1$, the coverage (fraction of length covered), $\theta(t)$, approaches a certain $k$-dependent jamming value $\theta(\infty) < 1$ for large times, $t$. This 1$d$ deposition model is in fact exactly solvable [3]. The jammed state is formed because there are gaps of lengths $b$, ..., $(k-1)b$, which cannot be filled by deposition. There exists a well defined continuum limit $k \to \infty$, for all quantities of interest.

However, for nonzero diffusion rate the coverage eventually reaches 100%, for large times, because diffusion allows the formation of large gaps within which a fraction of deposition attempts can succeed. Exact solution is no longer possible, even in 1$d$. Instead, the following line of argument has been advanced [4,5]. For large times, with $\theta$ near 1, most of the free length will be in isolated single-site gaps of length $b$. These can be brought together by diffusion to form large empty gaps, of at least $k$ lattice spacings, which can with some finite probability be covered by a depositing object (or be broken up again by diffusion). The asymptotics of the difference $(1 - \theta)$ for large times can then be related to the decrease in density of the diffusion-limited chemical reaction

$$kA \to \text{inert} , \tag{1.2}$$

where $A$ are diffusing reactants on a line, forming inert species, i.e., effectively annihilating, according to (1.2) with some finite probability on each encounter of $k$ reactants.

Such reactions have been studied in the literature [9]. The appropriate results will be reviewed later. For now it suffices to point out that for $k = 4, 5, \ldots$, and with some reservations (to be addressed later) for $k = 3$, one can use the mean-field rate equation which in terms of the coverage takes the form

$$\frac{d\theta}{dt} \simeq \gamma_k R\ell(1-\theta)^k . \tag{1.3}$$

While this approximate power-law behavior was confirmed by numerical simulations [5] for $k = 3, 4, 5, 10$, there is an important limitation to the rate equation (1.3): it has no obvious continuum limit as $k \to \infty$.

Indeed, no matter what $k$-dependence one conjectures for the dimensionless effective rate constant $\gamma_k$, the right-hand side of (1.3) cannot approach a definite function of $\theta$ as $k \to \infty$. This difficulty was also suggested by a preliminary observation based on the data for $k = 3, 4, 5, 10$, that the times $t$ from which (1.3) applies increase rapidly with $k$.

The main objective of the present work is to repair this deficiency of the mean-field theory in $1d$. Thus, in Section 2 we introduce a new approximation scheme, still mean-field in nature, which allows the derivation of improved rate equations which are valid in a significantly larger regime and which apply in all the limiting cases. Various implications of this new mean-field description will be detailed in Sections 3 and 4, with the latter devoted to the continuum limit. The original rate equation (1.3) turns out to hold only for

$$t \gg [(k-1)R\ell]^{-1} e^{k-1} \,, \tag{1.4}$$

which explains its failure to describe the continuum limit $k \to \infty$.

As with any mean-field approximation, fluctuations are expected to dominate the large-time behavior in some cases, such as for $k = 2$. The appropriate criteria are derived and discussed in Section 5. Finally, Section 6 is devoted to summary and discussion of those new aspects that must be considered in order to extend the theory to $d > 1$, which, however, is not attempted here; see earlier remarks in this section and the discussion in Section 6.

## 2. FAST DIFFUSION AND DIFFUSER SPACE

It is convenient to relate our system to a lattice model of hard-core particles (monomers) of length single lattice spacing $b$, each attempting to hop with rate $H = 4D/b^2$ per unit time to the left or right neighbor lattice site (rates $H/2$). We will use the term "diffusers" to refer to these particles. Hopping attempts succeed only if the hard-core constraint is not violated. The precise dynamics of diffusers is in fact not of interest to us here. The only property we will use is that for large times the system of hard-core particles approaches a steady state in which they are distributed statistically uniformly along the line [10].

Consider now a large length $L$ in our original deposition model. It is convenient to take $L$ to be the system size, in this section. We disregard finite-size effects by implying the thermodynamic limit first, in all the "intensive" expressions. The length $L$ contains $\theta L/\ell$ extended objects, each $k$ lattice spacings long. Disregarding deposition, diffusion in this system is the same as in a single-site-diffuser system in which all the extended objects of length $\ell = kb$ are replaced by diffuser particles of length $b$. The total length available to these diffusers is

$$\Lambda = (1-\theta)L + \theta L/k = \left(1 - \theta + \frac{1}{k}\theta\right) L \ . \tag{2.1}$$

The density of diffusers (per site of the diffuser-space lattice) is thus

$$\rho = \frac{\theta L/\ell}{\Lambda/b} = \frac{\frac{1}{k}\theta}{1 - \theta + \frac{1}{k}\theta} \ . \tag{2.2}$$

The gap size between every pair of neighboring diffusers can be $mb$, where $m = 0, 1, \ldots$. Note that $m = 0$ corresponds to contact, while the upper limit is taken to be infinity due to our disregard of finite-$L$ effects. Diffusion makes particle locations uncorrelated. In terms of the gap sizes, the appropriate statement is that the (normalized) probability to find gap size $m$ is

$$\text{Prob}\,(m) = \rho(1-\rho)^m \ . \tag{2.3}$$

Of course, the result (2.3) applies only for large times. However, we are proposing to use (2.3) as the approximate, fast-diffusion, mean-field form of the gap size distribution for the actual deposition model. The implications of this approximate assumption will be worked out in this and the next two sections. We will then proceed to derive the conditions for its asymptotic validity, in Section 5. Let us point out, however, that in the opposite extreme of no diffusion at all, exact results for the gap size distribution can be derived [3] although they were analyzed in detail only in the continuum limit $k \to \infty$. These results will be described in Section 4.

We now return to the original full-line problem, and assume that (approximately) the gap sizes are distributed according to (2.3), where the time-dependence enters via the $\theta$-dependence implied by (2.2). The total number of gaps (including $m = 0$) is equal to the number of objects. However, only gaps of size $m \geq k$ offer $(m - k + 1)$ distinct locations for deposition. Since the deposition attempt rate per lattice site is $Rb$, the overall average rate of successful depositions in $L$, per unit time, is

$$\omega = (Rb)(\theta L/\ell) \sum_{m=k}^{\infty} (m - k + 1)\rho(1 - \rho)^m = \frac{LR\theta(1 - \rho)^k}{k\rho} \, . \tag{2.4}$$

In each deposition, the coverage is incremented by $\ell/L$. Thus, we obtain the relation

$$\frac{d\theta}{dt} = \frac{\ell R\theta(1 - \rho)^k}{k\rho} \, . \tag{2.5}$$

Finally, we use (2.2) and introduce the dimensionless time variable

$$T \equiv R\ell t \, , \tag{2.6}$$

to write the equation for the coverage in the fast-diffusion approximation,

$$\frac{d\theta}{dT} = \frac{(1 - \theta)^k}{\left(1 - \theta + \frac{1}{k}\theta\right)^{k-1}} \, . \tag{2.7}$$

## 3. LATTICE DEPOSITION WITH FAST DIFFUSION

In this section we consider implications of the relation (2.7) for fixed $k$, i.e., for lattice deposition. Some conclusions are valid also in the limit $k \to \infty$ which, however, will be considered separately in the next section. We assume $k = 2, 3, 4, \ldots$ since the case $k = 1$ is trivial. In fact, (2.7) is exact for $k = 1$.

The curve $\theta(T)$ obtained by integrating (2.7) with the initial condition $\theta(0) = 0$ provides the upper bound (corresponding to $D = \infty$) for any deposition process with fixed diffusion rate (recall that the deposition rate was absorbed in the time scale); see (2.6). The exact solution for $D = 0$, available as quadrature [3], provides the lower bound. The two curves are in fact quite close for small and intermediate times as illustrated in Figure 1 for $k = 5$.

For large times, however, the behavior is quite different. The $D = 0$ curve approaches the jamming value $< 1$, at the exponential rate, $k^{-1}e^{-T/k}$, see [11]. However, the fast-diffusion result approaches the full 100% coverage, according to

$$1 - \theta(T) \simeq k^{-1}\left[(k-1)T\right]^{-1/(k-1)} . \qquad (3.1)$$

Interestingly, this power-law behavior sets in, and (2.7) can be replaced by the "reaction" type rate equation (1.3) with $\gamma_k = k^{k-1}$, for times

$$T \gg (k-1)^{-1}e^{k-1} , \qquad (3.2)$$

which become quite large as $k$ increases. This time scale is essentially determined by the condition $(1 - \theta) \ll k^{-1}$ imposed to have no variation due to the denominator in (2.7), for $\theta$ near 1. However, from (2.2) one can easily check that the same condition is also obtained from $(1 - \rho) \ll 1$, within the fast-diffusion formulation. The latter condition ensures single-lattice-spacing gap dominance, as discussed in Section 1, provided that the gap size distribution is approximated by (2.3).

The full result (2.7) is, however, more general than the rate equation (1.3); its precise limits of applicability will be derived in Section 5. As found for chemical reactions [9], and more generally for critical phenomena, the mean-field results can be used in those cases when local fluctuations do not dominate the dynamics.

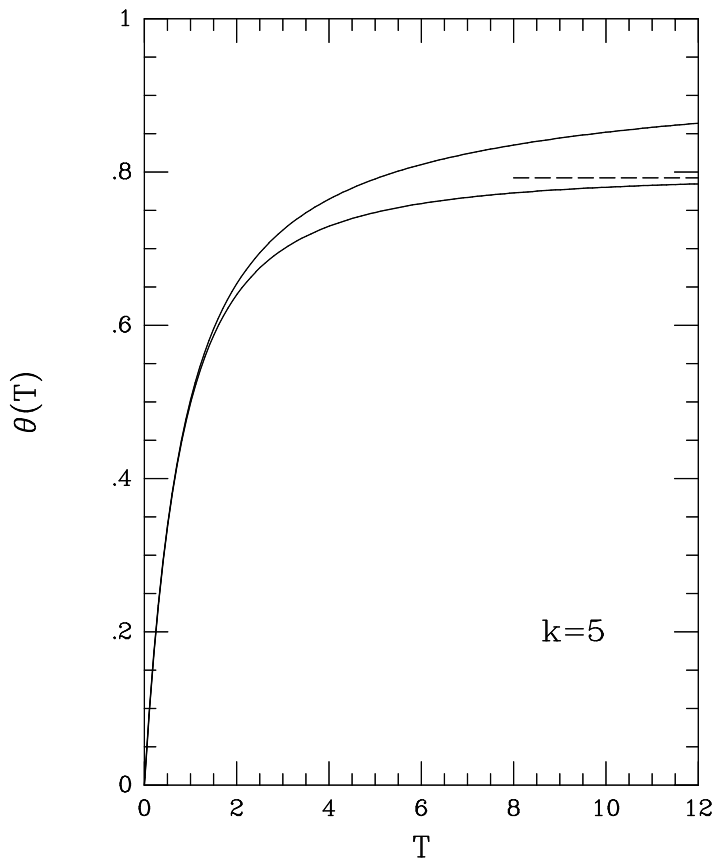

**Fig. 1.** The $k = 5$ curves $\theta(T)$ for $D = \infty$ (upper) and $D = 0$ (lower). The broken line marks the limiting, jamming value of the $D = 0$ coverage for large times. The $D = \infty$ coverage curve approaches 1 as $T \to \infty$; see Section 3.

However, even for the description of dynamics dominated by the uniform fluctuations describable asymptotically by effective-field type approaches some caution is called for. Indeed, one can claim that relation (2.7) provides the correct functional form of the asymptotic large time behavior, with the "large times" defined much less restrictively than when using (1.3); see Section 5. However, for any finite $D$ (and $k > 2$) to actually fit the data to the mean-field form, we must use it with "effective," renormalized rather than "bare" parameters. In (2.7) the only parameter was the overall rate. Phenomenological fits of data obtained for $D < \infty$, should be done with the right-hand side multiplied by an additional adjustable rate parameter of order 1, before integrating (2.7). This parameter will approach 1, and the time range for which (2.7) applies will extend down to $T = 0$, only in the limit $D \to \infty$.

Due to the limited numerical data available in the literature on the deposition problem [5] we have not attempted systematic tests of the above expectations. Some preliminary checks were made for $k = 10$. However, in a forthcoming publication [12] we report a similar improved mean-field approach formulated for the chemical reactions (1.2), as well as detailed numerical studies of the applicability of such "fast-diffusion" mean-field theories in $1d$.

## 4. CONTINUUM DEPOSITION

The $k \to \infty$ limit of (2.7) is easy to work out,

$$\frac{d\theta}{dT} = (1 - \theta) \exp\left(-\frac{\theta}{1-\theta}\right) . \qquad (4.1)$$

Thus the fast-diffusion approximation is well-defined. The approach to the limiting coverage is quite slow:

$$\theta \simeq 1 - \frac{1}{\ln(T \ln T)}, \qquad (T \gg 1) . \qquad (4.2)$$

This behavior is quite different from the other extreme of no diffusion at all. The coverage in the latter case reaches the jamming value ($<1$) at the rate $\sim T^{-1}$; see [11] for details.

The diffuser space concept was useful in our derivation, Section 2. However, it cannot be used directly in the continuum because the lattice spacing $b$ vanishes. It is useful to describe how the gap size probability in the fast diffusion approximation can be analyzed directly in the original space. We will also compare the fast-diffusion ($D = \infty$) expression to the result in the $D = 0$ case. (The main conclusions also apply for $k < \infty$. As already mentioned, the discussion is presented in this section because the $D = 0$ gap size distribution has been studied in detail only in the continuum limit [3,13].) The diffuser space is nevertheless an important tool in the derivation of the limits of validity of the mean-field approximation; see Section 5.

To analyze the gap sizes in continuum directly, in the fast-diffusion approximation, we note that for uncorrelated object locations, the gaps must be exponentially distributed. (As we just pointed out a similar line of argument can be carried out in the discrete case.) Let $g \geq 0$ denote the gap size (formerly $mb$), and $G(g)$ the density of gaps of size $g$ to $g + dg$ per unit length of the system, so that $G$ has units of $1/(\text{length})^2$. The decay width and the coefficient of the exponential can be fixed from the condition that the total number of gaps be equal that of objects, which can be reduced to

$$\int_0^\infty G(g)dg = \theta/\ell , \qquad (4.3)$$

and another condition: that the total length of all gaps be fraction $(1-\theta)$ of the system length,

$$\int_0^\infty gG(g)dg = 1 - \theta . \qquad (4.4)$$

The resulting exponential distribution function,

$$G(g) = \frac{\theta^2}{(1-\theta)\ell^2} \exp\left[-\frac{\theta g}{(1-\theta)\ell}\right], \qquad (4.5)$$

can be also shown to represent the correct $k \to \infty$ limit of the gap probability derived in Section 2 by "diffuser" arguments.

For $T \to \infty$ the exponential (4.5) collapses to the point $g = 0$. This is in contrast to the $D = 0$ distribution which only develops [3,13] a weak (logarithmic) integrable singularity at $g = 0$, and is finite up to $g = \ell$. For $g > \ell$ it is zero, which is just the manifestation of jamming: large gaps are filled up but the small ones are left over. For finite times, the singularity at the origin is not sharp [3,13]. There is a discontinuity in slope at $g = \ell$, and in fact for $g > \ell$ the functional form is exponential in $g$. Specifically, for large times, on approach to jamming, the $D = 0$ distribution follows $\sim \exp\left[-(T/\ell)g\right]$, whereas the fast-diffusion exponential (4.5) behaves for large times as $\sim \exp\left[-(\ln T/\ell)g\right]$.

The main effect of diffusion is therefore to increase the weight of large gaps at the expense of small gaps, which is particularly effective in the late stages of the deposition. These considerations actually suggest possible directions of improving the fast-diffusion approximation. A natural theoretical development, not attempted here, would be to write kinetic equations for the gap size distribution, incorporating the time-variation due to both diffusion and deposition.

## 5. CRITERIA FOR APPLICABILITY OF THE MEAN-FIELD THEORY

In order for the mean-field type approximations to hold, the system behavior must be dominated by fluctuations in various overall averaged quantities. If instead local density fluctuations dominate the deposition process, then mean-field theory will not be applicable. Thus, for mean-field theory to apply, density fluctuations due to successful deposition attempts must be effectively smoothed out by diffusion, over a certain characteristic length scale $L$ (which in this section no longer denotes the system size).

For short times the substrate is nearly empty, the deposited objects are uncorrelated and so $L$ will be small. We expect $L$ to become large for large times. In order to estimate $L$, we make two observations. Firstly, the average distance between two neighboring empty sites is

$$\Delta L = \frac{\theta \ell}{k(1-\theta)} \; . \tag{5.1}$$

This distance becomes large for $\theta$ near 1. Secondly, each successful deposition eliminates $k$ empty sites, which were at that instant part of a gap of $m \geq k$ sites. The overall gap distribution will thus be appreciably "perturbed" over a length

$$L = O(k)\Delta L \; , \tag{5.2}$$

which represents the average size of the region which would have been occupied by $O(k)$ empty sites after diffusive spread, had there been no deposition event.

In order to have a more definite expression, we note that for $k = 1$ there are no correlations at all, and formally, $L = 0$. Thus we replace the $O(k)$ factor by $(k-1)$,

$$L = \frac{(k-1)\theta \ell}{k(1-\theta)} \; . \tag{5.3}$$

This prefactor choice is, however, purely phenomenological.

For any region of length $L$, the rate $\omega$ of successful deposition events in that region is given by (2.4). The time scale it takes the diffusion to equilibrate density perturbations in length $L$ is

$$\tau = \Lambda^2/D ,  \quad (5.4)$$

where $\Lambda$ is the diffuser-space equivalent of $L$, see (2.1).

The condition to have effective diffusional equilibration is

$$\omega\tau \ll 1 , \quad (5.5)$$

which upon collecting all the relations and definitions introduced above reduces after some algebra to

$$\frac{D}{R\ell^3} \gg \left(\frac{k-1}{k}\right)^3 \theta^3 \left(\frac{1-\theta}{1-\theta+\frac{1}{k}\theta}\right)^{k-3} . \quad (5.6)$$

Note that we used relation (2.2) to replace $\rho$-dependence by $\theta$-dependence.

The right-hand side of (5.6) is shown in Figure 2 for $k = 2, 3, 4, 6$ and $\infty$, where in the limit $k \to \infty$ we have

$$\frac{D}{R\ell^3} \gg \theta^3 \exp\left(-\frac{\theta}{1-\theta}\right) , \qquad (k = \infty) . \quad (5.7)$$

The main conclusion is that for $k = 4, 5, \ldots, \infty$ the fast-diffusion mean-field theory should be asymptotically correct for large times. Successful depositions are so far and few between, that the system has enough time to equilibrate diffusively between depositions. The point brought up in Section 3, regarding the use of the effective "renormalized" rate to actually fit the data by the mean-field functional forms, must be kept in mind, however. The case $k = 3$ is borderline. Our estimates "from within mean-field" cannot be used to decide if fluctuations will be marginally relevant in this case, i.e., if additional logarithmic factors will make the appropriately modified version of (5.6) break down at large times. Phenomenologically, however, the mean-field expressions with "effective" rate constant may provide a good quantitative fit of any data set taken over a large but finite number of decades in $T$.

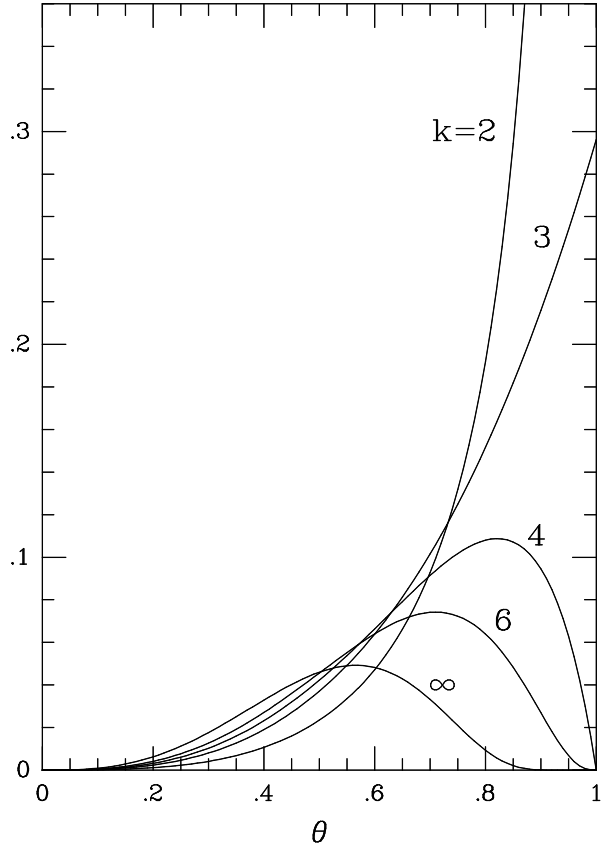

**Fig. 2.** The functions defined by the right-hand sides of (5.6), for $k = 2, 3, 4, 6$, and (5.7), for $k = \infty$.

For $k = 2$ the mean-field theory breaks down similarly to chemical reactions [9]. Other approaches which account for local fluctuations must be adopted; see [4] for such developments and numerical tests.

## 6. DISCUSSION

In this last section we begin by addressing the issue of how various ingredients of the theory are modified for $d > 1$. Firstly, fluctuations are always smaller in higher dimensions and therefore mean-field theories would provide a useful tool. However, as mentioned in Section 1, there are several points of difficulty. Locally, in higher dimensions there can be immobile "gridlocked" gaps (for nonspherical object shapes or on lattices), and it is not even obvious if the large time coverage will always reach the maximal, fully crystalline packing value. In fact, the final state may be amorphous, or polycrystalline with a network of defects.

Indeed, polycrystalline and amorphous structures were found in numerical experiments on the formation of packed structures in two and three dimensions [14] by methods other than random sequential filling with diffusional relaxation. (We are not aware of any numerical investigations of the latter process in $d > 1$.) In fact, the tendency was [14] for polycrystallinity in $2d$ and amorphous structures in $3d$.

Globally, neither the dynamics of hard-core diffusing extended objects nor the collective aspects of the equilibrium state are well understood, even without added deposition [8]. Glassy slowing down and phase transitions are among the collective effects possible. Thus, even within the fast diffusion approximation, the problem is much more complicated than in one dimension.

Another aspect of the difficulty in extending the theory to $d > 1$ is that numerical data are quite difficult to collect, and the currently available experimental results [6] are few and very recent. However, the complexity of the problem also suggests that new interesting effects may be discovered. As a first step, heavy numerical effort is called for.

In the present work, we accomplished the formulation of the mean-field approach directly for the deposition problem, without having to map to the chemical reactions. The main improvement was the well defined large-$k$ behavior of the new mean-field expressions, and a larger regime of validity for each value of $k > 3$. The criteria of applicability of the mean-field theory were derived in the deposition-model context. A new finding was that mean-field theory provides the correct asymptotic description of continuum deposition, with diffusional relaxation, on a line.

This research was partially supported by the Science and Engineering Research Council (UK) under grant number GR/G02741. One of the authors (V.P.) also wishes to acknowledge the award of a Guest Research Fellowship at Oxford from the Royal Society.